\documentclass[11pt]{article}
\usepackage[letterpaper, hmargin=1in, vmargin=1.2in]{geometry}
\usepackage{color}

\usepackage{amssymb}

\usepackage{cite}

	\usepackage[pdftex]{graphicx}
	\graphicspath{{./figs/}}
  	\DeclareGraphicsExtensions{.pdf,.jpeg,.png}
\usepackage[cmex10]{amsmath}
\usepackage{amsfonts}
\usepackage{algorithm}
\usepackage{algpseudocode}

\usepackage{array}

  \usepackage[caption=false,font=footnotesize]{subfig}
\usepackage{fixltx2e}
\usepackage[colorinlistoftodos]{todonotes}
\usepackage{enumerate}

\usepackage{rotating}
\usepackage{multirow}

\usepackage{amsthm}

\usepackage[T1]{fontenc}
\usepackage[utf8]{inputenc}
\usepackage{authblk}

\usepackage[T1]{fontenc}
\usepackage{lmodern}

\usepackage{hyperref}
\hypersetup{  colorlinks, citecolor=green, linkcolor=blue, urlcolor=blue}

\begin{document}
%
\title{Intrinsically Reliable and Lightweight Physical Obfuscated Keys}

\author{Raihan Sayeed Khan}
\author{Nadim Kanan}
\author{Chenglu Jin}
\author{Jake Scoggin}
\author{Nafisa Noor}
\author{Sadid Muneer}
\author{Faruk Dirisaglik}
\author{Phuong Ha Nguyen}
\author{Helena Silva}
\author{Marten van Dijk}
\author{Ali Gokirmak}

\affil{University of Connecticut \\Email: raihan.khan@uconn.edu}



\maketitle



%

\begin{abstract}
Physical Obfuscated Keys (POKs) allow tamper-resistant storage of random keys based on physical disorder. 
The output bits of current POK designs need to be first corrected due to measurement noise and next de-correlated since the original output bits may not be i.i.d. (independent and identically distributed) and also public helper information for error correction necessarily correlates the corrected output bits.
For this reason, current designs include an interface for error correction and/or output reinforcement, and privacy amplification for compressing the corrected output to a uniform random bit string. We propose two intrinsically reliable POK designs with only XOR circuitry for privacy amplification (without need for reliability enhancement) by exploiting \textbf{variability of lithographic process and variability of granularity in phase change memory (PCM) materials}. The two designs are demonstrated through experiments and simulations. 
\end{abstract}



\newpage
\tableofcontents
\newpage

\section{Introduction}
\label{sec:intro}

Since \textbf{Physical Unclonable Functions} (PUFs)~\cite{Pappu2001} were introduced in 2001, they have been considered as promising hardware security primitives for building secure systems such as IP protection, identification and authentication protocols, cryptographic primitives, etc.~\cite{Kumar2008,Armknecht2009,Maes2010,Yu2011,Brzuska2011}. 

PUFs exhibit a random challenge-response behavior, i.e., a random-looking response is generated by a PUF when queried with a specific challenge based on intrinsic manufacturing variations. All PUF behaviors are unique and \textit{physically unclonable}.

We partition PUF designs into two different categories: \textit{strong PUFs} and \textit{weak PUFs} (also called Physical Obfuscated Keys or POKs)~\cite{Gassend2003a}. The major difference between these two categories is that a strong PUF has a large number of challenge-response pairs (CRPs), while a weak PUF or POK has a limited number of CRPs. 

In this paper we focus on POKs, which due to its important security properties such as physical unclonability with tamper-resistance, uniqueness,  and unpredictability (or randomness) can be used for IP protection, authentication protocols~\cite{Suh2007}, tamper-resistant storage of random keys or building digital PUFs~\cite{Gassend2003a}.
 For example, in the digital PUF concept a keyed Pseudo Random Function (PRF) is used as a PUF where challenges are inputs to the PRF and responses are the outputs of the PRF and where the key is a POK response of a fixed challenge to the POK.
%
%
Since there is no secret key stored in non-volatile memory (NVM) in a POK-based system, the adversary cannot directly retrieve the secret key from the device. 

Typically, the POK circuit consists of two parts: a randomness source and a randomness extraction circuit. The randomness extraction circuit extracts and converts the analog signal from a randomness source into a response in the form of  \textbf{digital raw output} to be used in further digital computation. For example, in a Ring Oscillator (RO) based POK~\cite{Suh2007}, the randomness source is the delay randomness of circuit components  and the randomness extraction circuit consists of RO pairs with counters and comparators. For each RO pair its frequencies (measured by a counter) are compared (by a comparator)  to produce one bit raw output. The collection of the digital raw output bits from all RO pairs together forms the response of the POK. In this paper we consider POKs with just {\em one} challenge response pair, i.e., when powered-up a POK just (re)generates one fixed random looking response (after randomness extraction).

One feature of a POK is that if an adversary physically opens the device for reverse engineering purpose, then the POK's randomness source and consequently raw output (i.e., response)  is changed. Therefore, if the POK's raw output is used as a secret key in a symmetric key system or as a key mask in public key systems, the adversary cannot physically open the device for retrieving such a secret key.  

A POK design must offer POKs with the following properties:
\begin{itemize}
\item \textbf{Robustness/Reliability:}  The POK response is stable (i.e., the POK is able to regenerate the same response over and over).
\item  \textbf{Unpredictability:} The POK response is not predictable (without access to the POK), i.e., it should be randomly generated.
\item \textbf{Uniqueness:} A POK's response is different from device to device.
\end{itemize}
These properties are necessary in order for a POK response to be used as a key  in a secure system. In particular, a POK's response bits must be reliable and independent and identically distributed (i.i.d.) (i.e. not correlated). Both of these required properties may not immediately be satisfied for basic POK designs (such as the afore mentioned RO POK): To overcome the reliability and unpredictability problem, an Error Correcting Code (ECC)~\cite{Gassend2002a,Suh2007} with Privacy Amplification (PA) or Fuzzy Extractor (FE)~\cite{dodis2004fuzzy} is exploited to make a POK's (or more generally a PUF's) output robust and unpredictable. However, this approach leads to additional requirements in terms of \textbf{hardware overhead and efficiency}  as discussed in detail in Section~\ref{subsec:relibility_enhancement}.

We notice an alternative approach for a POK/PUF design based on bi-stable memories (for example SRAM memory) for which  the reliability of the POK/PUF output can be significantly improved by output reinforcement processes as discussed in~\cite{BhargavaCakirMai2012,BhargavaMai2013}, for example, by the Hot Carrier Injection technique. Typically, this process changes the POK/PUF's circuit characteristics permanently and thus the POK/PUF's output can be highly reliable. To achieve this goal, however, a suitable setup is required.

In this paper we search for other alternative and practical POK designs that allow a (basic) randomness extraction circuit which does not need any ECC circuitry because the randomness source of the POK itself is intrinsically reliable such that a simple comparator based solution produces bits that are both reliable and as a collective have high min-entropy. Due to having a high min-entropy, the de-correlation problem (making the final response i.i.d.)  can be solved in a simple way by using the de-correlation technique from True Random Number Generator (TRNG) constructions, where typically several bits are XORed together resulting in i.i.d. response bits. This leads to lightweight POK designs without expensive ECC and PA circuitry. 

  
If it is possible to design \textit{intrinsically} reliable POKs based on common technology  without needing a specialized set-up per fabricated POK, then no reliability enhancement techniques are needed and resulting in lightweight POK designs as explained above. We propose two new intrinsically reliable POK designs:



\vspace{3mm}

\noindent
\textbf{POK based on the Variability of Lithographic Process:}
First, we observe that a lithographic process can provide a good random source if applied below the particular process limit. Typically, the lithographic process can be used to create a \textbf{Unique Object (UNO)}~\cite{Ruhrmair2011}. In~\cite{Ruhrmair2011}, a UNO is described as  ``A physical system that, upon measurement by an external apparatus, exhibits a small, fixed set of inimitable analog properties that are not similar to any other objects.'' A UNO is different from a POK in the way we measure the objects. For a POK, we have to use an internal apparatus so as to not expose the output while UNO's output is measured by an external aparatus and is not kept as a secret. As an important property, the output of a UNO is intrinsically reliable.


Typically, the smallest features on a chip are safely above the lithographic process limit to ensure a high yield. But features at this limit can be exploited to obtain a random array of functional/broken devices. Hence, the connectivity of each such line cell can be converted to a random bit. This means  a secret key or a key mask based on the connectivity of cells can be constructed in a robust way. 
%

Notice that a set of line cells constructed by a lithographic process is a UNO, however, if this set is hidden, then we obtain a POK instead of a UNO. 
Since the width of cells is very small, we can have many random bits giving a high density key. 


In PUF literature, the lithographic process has been introduced to construct PUF designs as shown in~\cite{SreedharKundu2011,KumarBurleson2013,KumarDhanuskodiKundu2014}, see Section \ref{subsec:lithoPUF}. However, our proposed design based on connectivity is very reliable and requires minimal randomness extraction circuitry.


\vspace{3mm}

\noindent
\textbf{POK based on the Variability of Granularity in Phase Change Memory:}
As a second contribution, we propose a novel Phase Change Memory (PCM) based POK. Existing PCM-based PUFs~\cite{ZhangKongChang2013,ZhangKongChangEtAl2014} compare resistances of two PCM cells which differ due to process variation in order  to produce one output bit. The resistances are not reliable due to resistance fluctuations over time. Our proposed PCM-based POK differs from those in~\cite{ZhangKongChang2013,ZhangKongChangEtAl2014} in terms of design concept. We use a single PCM device with three contacts to produce one bit based on \textbf{grain map variability in the PCM device}. Our proposed PCM cell randomly programs one of two contacts, thus a 0 or 1 is encoded. We can achieve a reliable PCM-based POK without using any reliability-enhancement techniques as mentioned above.  
 

\vspace{3mm}

\noindent
\textbf{Organization:}
We discuss related works in Section~\ref{sec:notation_background}. Section~\ref{sec:lithography} presents the lithography-based POK. The PCM-based POK is described in Section~\ref{sec:pcm}. Section~\ref{sec:conclusion} concludes this paper.

\section{Background and Related Work}
\label{sec:notation_background}

In this section, we provide background and related work on reliability enhancement techniques, lithography-based PUFs and PCM-based PUFs. 

\subsection{Reliability enhancement techniques}
\label{subsec:relibility_enhancement}




In general, there are two approaches to make a POK reliable: Response Reinforcement and  Error Correction.


\vspace{3mm}

\noindent
\textbf{Response Reinforcement:} Response reinforcement  upgrades a PUF's reliability by using normal detrimental IC aging effects. Bhargava et al. \cite{BhargavaMai2013} demonstrate a one-time post-manufacture step of hot carrier injection stress in one of the NMOS transistors in a sense amplifier. The enhanced offset magnitude of the sense amplifier for both positive and negative polarities eliminates chances of errors due to environmental (voltage and temperature) variations and aging. Additional error correction circuit overhead and possible  information leakage  through helper data, see below, are alleviated in this approach. Even though 1.7 years of reliable operation is simulated by bake experiment, the deleterious long stress time of 125s for a feature size of 65 nm might affect reliability at matured stages. ~

\begin{figure}[!t]
\centering
\includegraphics[width = 0.6\columnwidth]{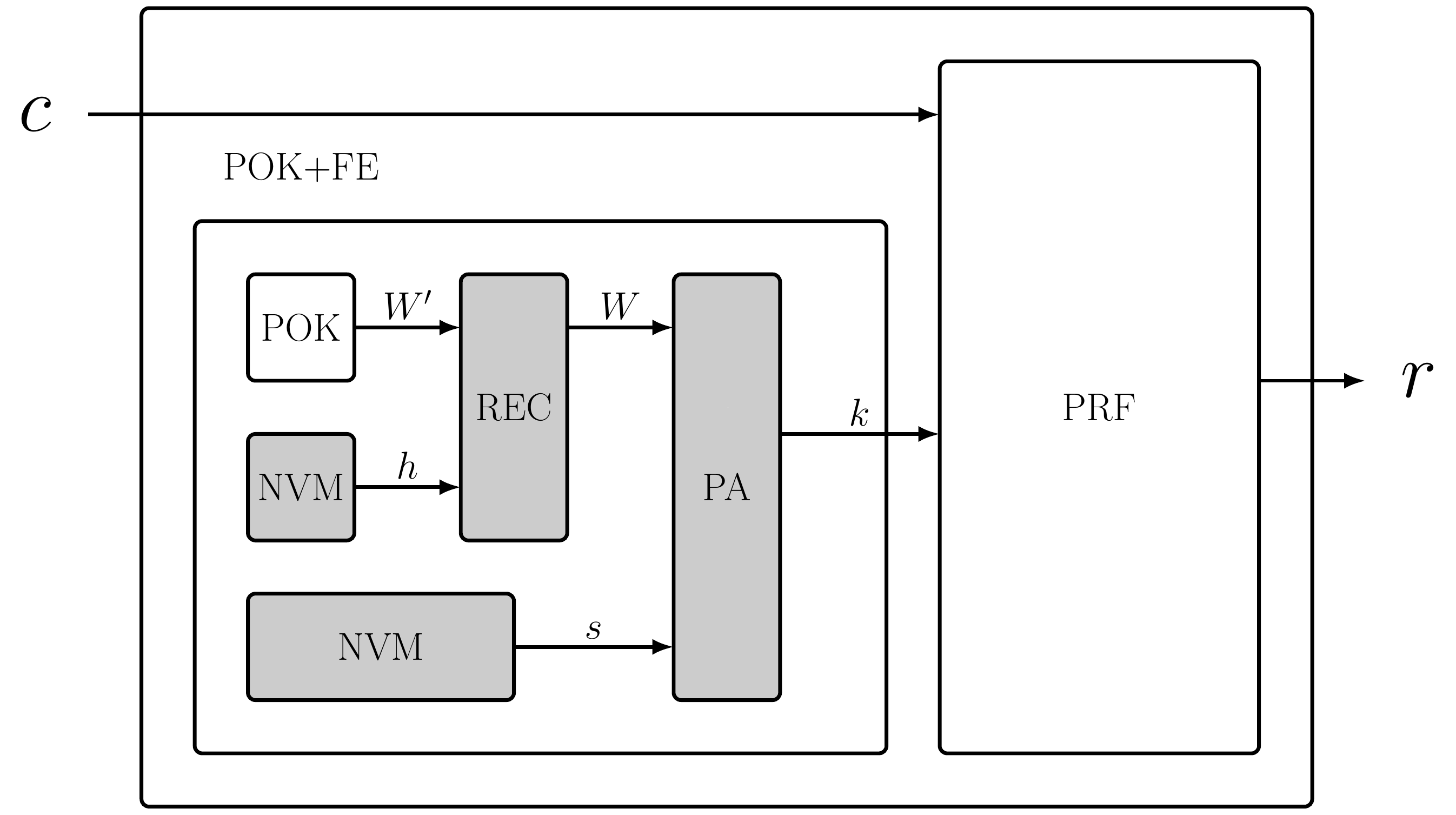}
\caption{A Digital PUF based on a POK architecture where POK represents the basic POK enhanced with a Fuzzy Extractor (FE) in Rep mode (using a RECovery procedure and Privacy Amplification or PA)  and extended with a Pseudo Random Function (PRF). 
%
} 
\vspace{-1 em}
\label{fig:DigitalPUF}
\end{figure}


\vspace{3mm}

\noindent
\textbf{Error Correction:} Compared to response reinforcement,  error correction does not change the behavior of the output permanently. An extra circuit implements an error correction algorithm. Several constructions have been proposed: Fuzzy Extractors (FEs)~\cite{dodis2004fuzzy}, Computational Fuzzy Extractors~\cite{FullerMengReyzin2013}, and LPN-based PUFs~\cite{HerderRenDijkEtAl2016}. We notice that there are many different ECC-based constructions as shown in~\cite{Delvaux2014,Nguyen2015}. 
 Without loss of generality in terms of our discussion regarding hardware overhead and efficiency, we focus on FE~\cite{dodis2004fuzzy}. 

FE has two procedures: $\mathrm{Gen}$ and $\mathrm{Rep}$, the latter depicted in Figure~\ref{fig:DigitalPUF}. $\mathrm{Gen}$ takes basic POK output $W$  and a true random string $s$ as inputs and produces output $(h,s)$ where $h$ represents helper data. 
  Helper data $h$ and random string $s$ are programmed in the POK architecture and used to reconstruct $W$ and a derived key $k$ from a noisy version $W'$ generated as a result of a next measurement of the noisy source in the Rep procedure. $\mathrm{Gen}$ generates helper data (i.e., parity check information) $h$ for $W$ using an error-correcting code in, what is called, a \textit{Secure Sketch} (SS). SS uses a random bit vector $x$ for this purpose, i.e., $h\leftarrow \mathrm{SS}(W,x)$. 


Figure~\ref{fig:DigitalPUF} depicts the Rep procedure of a FE~\cite{dodis2004fuzzy} in a POK architecture.
Rep takes as input a next measurement $W'$ of basic POK output. $W'$ is a noisy version of $W$. If $W'$ and $W$ are different in $\leq d$ positions (i.e., the Hamming distance $d_{\mathrm{H}}(W',W)\leq d$), then the helper data $h$ from the Gen procedure allows a Recovery  procedure (REC) to reconstruct $W$ from $W'$. In fact, REC can be thought of as the reverse algorithm of SS. 
Next a random key $k$ is generated by applying a de-correlation process, called \textit{Privacy Amplification} ($\mathrm{PA}$),  to $W$. This is needed because the POK may produce correlated bits, but more important, within the attacker's view the helper data $h$ (and random string $s$) can easily be read because $h$ (and $s$) is either stored in NVM or hard coded by fuses, and this means $h$ is known to the adversary and conditioned on $h$ the reconstructed bits in $W$ are correlated. PA computes  $k \leftarrow \mathrm{PA}(W,s)$ and involves a hash function~\cite{dodis2004fuzzy}.  Since $W$ is reliably reconstructed, $k$ is also reliably reconstructed  and due to PA, $k$ is a random secret key. In Figure~\ref{fig:DigitalPUF}  $k$ is used for a digital PUF which if taking a challenge $c$ as input, produces an output response $r=\mathrm{PRF}_k(c)$ where PRF is a pseudo random function.




Error correction based approaches have the following limitations:

\begin{itemize}
 

\item \textit{Hardware Overhead.} As pictured in Figure~\ref{fig:DigitalPUF}, besides the basic POK circuit, several additional digital circuits (depicted in grey) need to be implemented and this accounts for a large hardware overhead of the final POK architecture. 


\item \textit{Efficiency.} Since many digital processes are involved in producing a secret key $k$, the efficiency of POK-based systems is  significantly reduced in terms of execution time and power consumption (see~\cite{BhargavaCakirMai2012,BhargavaMai2013}). 

\end{itemize}

The limitations discussed above can be avoided if the basic POK output bits are intrinsically reliable: All the grey-color blocks in Figure~\ref{fig:DigitalPUF} can be removed. \textit{Response reinforcement}  can also avoid these limitations but it is only applicable to bi-stable memory and requires an additional process per POK.

\vspace{3mm}

\noindent
\textbf{De-correlation.} Even if we have an intrinsically reliable basic POK, its  output bits are likely correlated. Thus, we cannot directly use these bits as a secret key. We first need to de-correlate these bits by a privacy amplification process as described in the error correction based approach above. If PA is required, then the hardware overhead of the system is increased and the efficiency of system is reduced.   
However,
if the to-be-decorrelated bits are reliable and in addition have high min-entropy, a very simple trick used in the construction of True Random Number Generators applies: We can perform an XOR operation on several correlated bits together to produce a response bit. This results in close to independent and identically distributed response bits. 

\subsection{Lithography based PUFs}
\label{subsec:lithoPUF}


Using optical lithography, devices and interconnect layers are printed layer by layer in specified patterns on semiconductor wafers using a monochromatic light source focused through a lens and mask system. The entire IC fabrication process consists of hundreds of patterning, deposition, etching, doping, and polishing steps, all with inherent variabilities~\cite{Mack2007,KunduSreedharSanyal2007,KumarDhanuskodiKundu2014,SreedharKundu2011}. These include irregularities in etching, doping and polishing steps, light source intensity fluctuations, lens and mask defects and alignment, defocus, and interference problems~\cite{KumarDhanuskodiKundu2014}. The variations in the final dimensions of a device are a combination of variations introduced by the lithographic exposure, photo-resist processing and etching steps. For typical logic and memory applications, these variations must be within tolerances required for the desired yield and reliability. For other applications however, such as security, these variations may be utilized to achieve inherent, true physical randomness. Process variations can be systematic, such as mask defects or deviations from layout design, which are reflected similarly on all dies, or ``random'', such as local fluctuations in exposure, etch, deposition or polishing steps, which result in device-to-device variations within each die or die-to-die variation.

Several resolution enhancement techniques, such as optical proximity correction, alternating phase shift mask, and immersion lithography, which are used to overcome sub-wavelength lithographic challenges to print  $\approx 10$ nm features  using a much longer wavelength UV light source (193 nm)~\cite{KunduSreedharSanyal2007,KumarDhanuskodiKundu2014}, are intentionally avoided to engineer highly sensitive lithographic variations based PUFs (litho-PUFs)~\cite{KumarDhanuskodiKundu2014,SreedharKundu2011}. Exploiting linewidth, length, and height variations in lithography simulation of litho-PUFs with forbidden pitch, Sreedhar et al. demonstrated output voltage fluctuations as a function of focus variation in the simulated devices~\cite{SreedharKundu2011}. Kumar et al. addressed fluctuations in the light intensity and durations and focus variations due to wafer tilt and resist thickness variations and showed improved inter-die and inter-wafer uniqueness with lithography simulation in the forbidden pitch zone~\cite{KumarDhanuskodiKundu2014}. To improve the performance of litho-PUFs the systematic variations should be suppressed and the random variations preserved or enhanced \cite{KumarDhanuskodiKundu2014}. Forte et al. delineated OPC optimized for litho-PUF which enhances the random variations while reducing the systemic variations \cite{Forte2012a,Forte2012}.

Wang et al. proposed a stability guaranteed PUF based on random assembly errors which result in random permanent connections during a directed self assembly (DSA) process \cite{Puneet}. However, the results presented in \cite{Puneet} were obtained from simulations and did not contain hardware implementation data. The proposed PUF is also susceptible to invasive attacks.


%

\subsection{PCM-based PUFs}


Phase change memory (PCM) is an emerging non-volatile computer memory technology \cite{WongRaouxKimEtAl2010}. It offers high speed, high endurance, and non-volatility. A typical PCM cell consists of a phase change material (typically $Ge_2Sb_2Te_5$ or GST) between two electrodes. A small volume of this material in between the two contacts can be changed quickly and reversibly between the amorphous phases and the crystalline phase - leading to a high resistance state (Reset) and low resistance state (Set) - by suitable electrical pulses that result in different self-heating. Amorphization is achieved by melting followed by abrupt cooling (melt-quench) whereas crystallization is achieved by maintaining the element above crystallization temperature for a sufficiently long period of time, either by a longer, lower voltage pulse or by a melting pulse followed by a longer fall time. The read operation is performed with a small voltage pulse such as not to disturb the crystalline state of the cell~\cite{WongRaouxKimEtAl2010,burr2010phase}. 
The process variations and programming sensitivity of PCM devices have been utilized by Zhang et al. for reconfigurable PUF , that dynamically generates updatable cryptographic keys. The address of two selected PCM cells is used as the challenge and the resistance comparison between the two cells is the response~\cite{ZhangKongChangEtAl2014}. 


%
%
%

\section{Physical Obfuscated Key Based on Lithography Limit}
\label{sec:lithography}

\begin{figure}[!t]
\centering
\includegraphics[width = 0.4\columnwidth ]{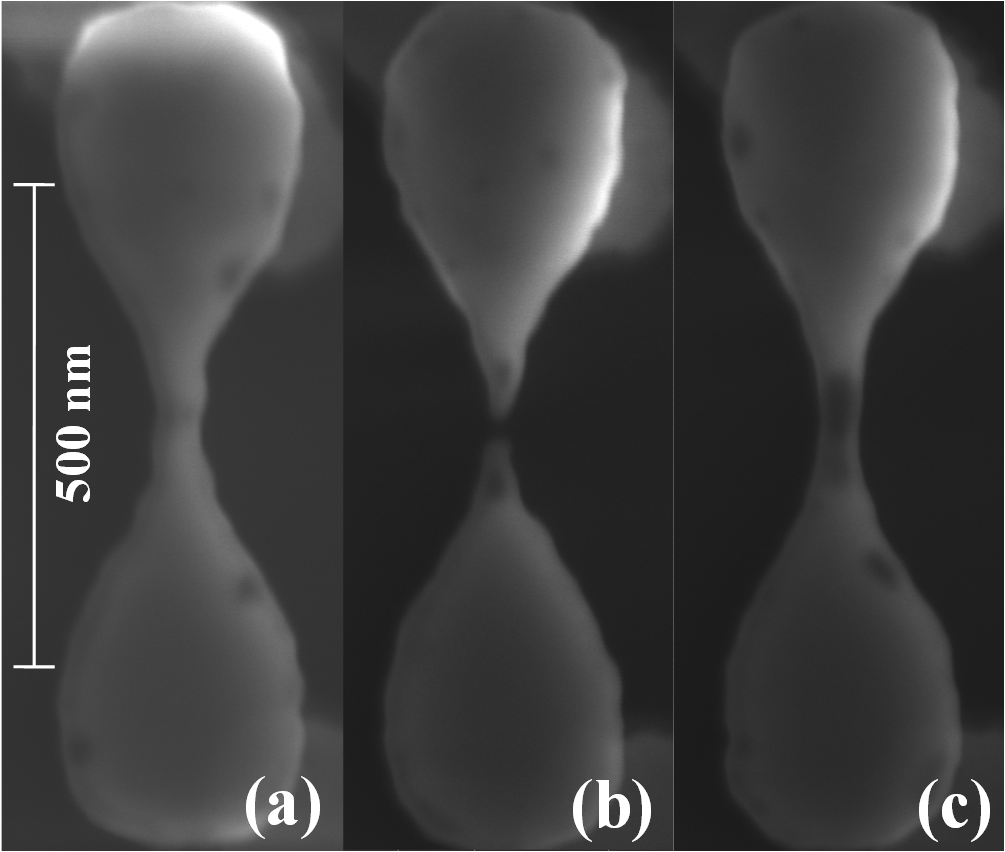}

\caption{SEM image of (a) connected cell, (b) broken cell, (c) cell with a void formed at the center.}
\vspace{-1 em}
\label{fig:sem_cell}
\end{figure}

\begin{figure}[!t]
\centering
\includegraphics[width = 0.8\columnwidth]{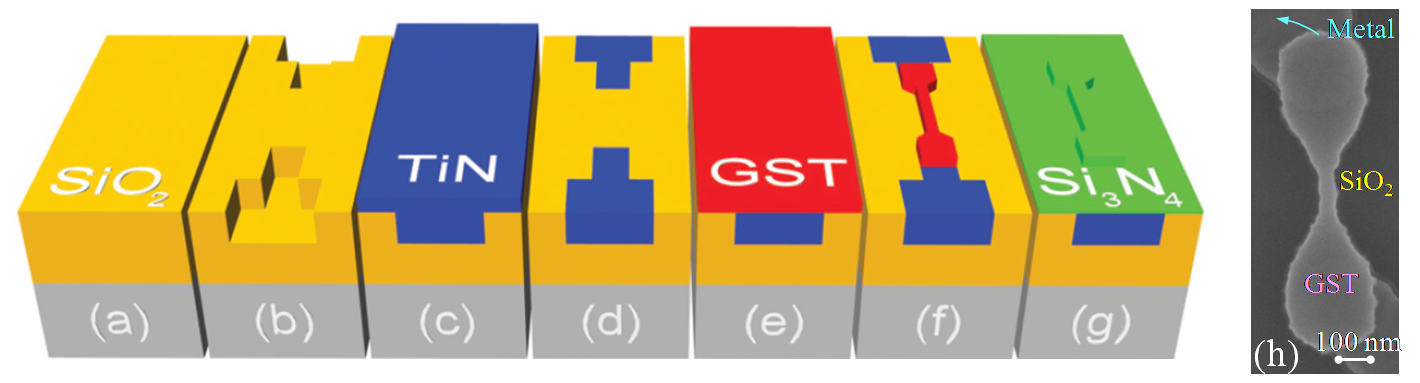}
\caption{Illustration of GST line cell fabrication steps (a) 700 nm $SiO_2$ growth, (b) 250 nm deep trench formation, (c) 300 nm TiN fill, (d) chemical mechanical planarization (CMP), (e) GST film deposition (50 nm), (f) patterning of GST film, (g) $Si_3N_4$ cap layer deposition, (h) SEM image of a fabricated line cell~\cite{DirisaglikBakanJuradoEtAl2015}. }
\vspace{-1 em}
\label{fig:fab_cell}
\end{figure}\

As shown in~\cite{BhargavaMai2013}, a reliable SRAM POK can be made by applying a response reinforcement technique. However, to produce one raw bit secret,  6 transistors are needed, i.e., one SRAM cell. Moreover, SRAM POK is feasible only when the device has SRAM as memory. Our approach can produce one raw bit secret at the cost of 1 transistor and one line cell (very small compared to the transistor), and our POK can be implemented in \textbf{any} device.

For high-yield, reliable applications the smallest dimensions should be safely above the lithographic resolution limit at which some devices are successfully printed and others are not (Figure~\ref{fig:sem_cell}). For a given lithography process, the yield can be varied between 0\% and 100\% for devices with sizes well below the resolution limit and 100\% for devices with sizes well above the resolution limit. In this work we propose to use `lithography-limit' devices to implement true, physical randomness as connected or disconnected lines in a dense array of two contact devices. In this work, GST line cells are used to demonstrate the proof of concept, but the same idea can be applied to other materials (e.g. silicon, metal) to produce  a reliable POK.

\begin{figure}[!t]
\centering
\includegraphics[width = \columnwidth]{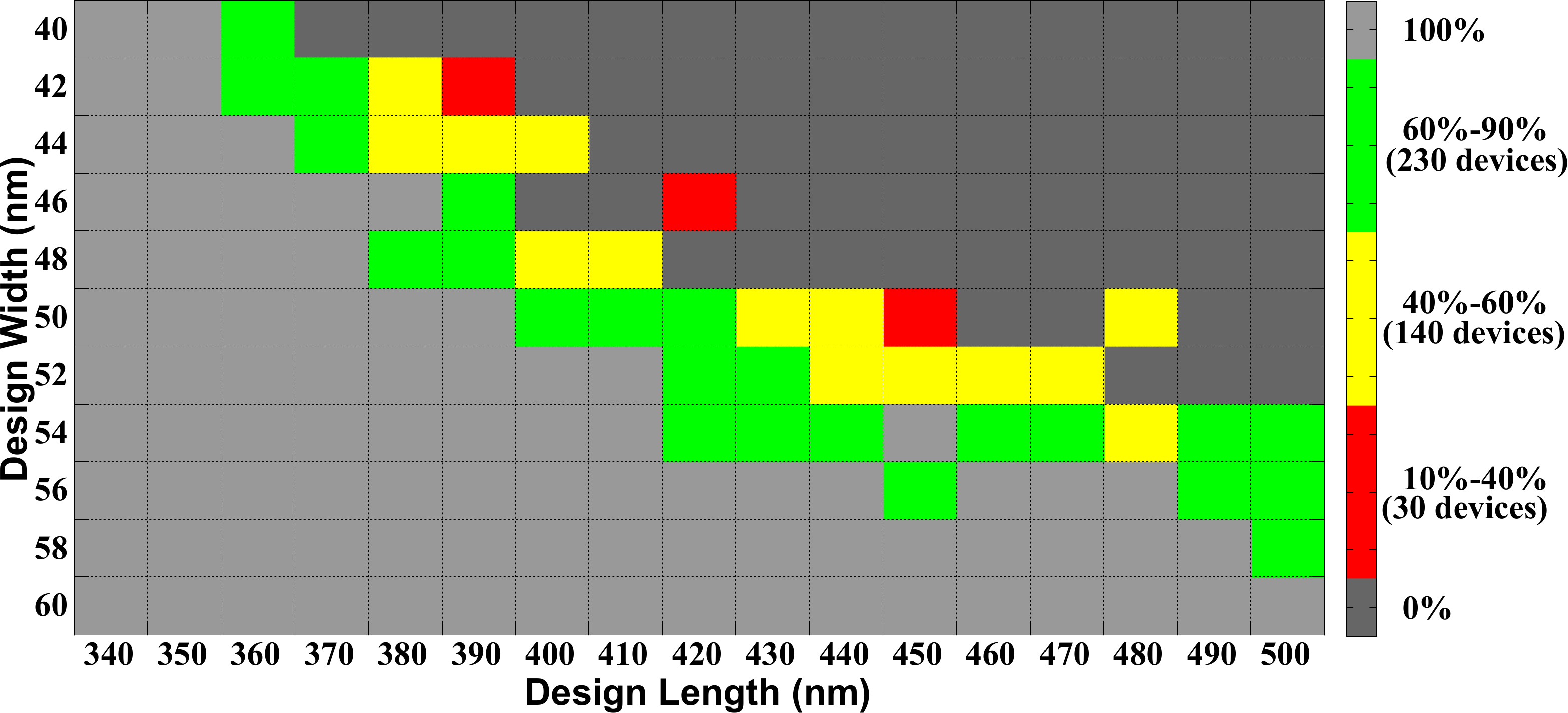}
\caption{{Connectivity yield color map for two-contact line cells  with widths below the lithography limit (90 nm technology) for 10 dies ($\thicksim$1870 devices). Each block represents a group of line cells with similar design length and width. The width of the cells varies from group to group by 2nm (y axis) and length varies by 10 nm (x axis). The colors represent percentage of line cells that are functional in a group consisting of 10 cells of same design dimensions (e.g. yellow corresponds to size blocks where 4 to 6 out of the 10 cells are working/connected). The 40-60\% region of cell dimensions can be used to generate random, unique security keys. A yield map needs to be obtained for a given fabrication technology process which will vary depending on fabrication parameters.}}
\vspace{-1 em}
\label{fig:litho_width}
\end{figure}

The bottom contacted 50 nm thick GST line cells used in the experiments were fabricated on 700 nm  silicon dioxide ($SiO_2$) thermally grown on Silicon wafers as described in  \cite{DirisaglikBakanJuradoEtAl2015}. Devices with varying widths and lengths were fabricated using 90 nm technology. Design widths start from 40 nm, well below the expected capability of  lithography, and increase with 2 nm increments. We observe that at small device lengths the wider contact regions merge, hence GST structure is continuous from one contact region to the other, even for 40 nm design widths, while the narrow devices do not survive the pattern transfer process at longer design lengths. Hence, connectivity between the two contacts of a device is random for the cells that are long enough and the widths are at the lithography limit, e.g. cells with $L_{design} = 460 nm$, $W_{design} = 52 nm$ display 40\%-60\% probability of having electrical connectivity measured in 10 dies\footnote{10 dies from a single 200 mm wafer were used to demonstrate the proof of concept.} (Figure~\ref{fig:litho_width}). 

Cells within this range can be used as bits in a string specific to the device, with a functional and broken cell representing 1 and 0 respectively. That way each chip will have a secret key of its own, thereby forming a physical obfuscated key. The output from the proposed POK is reliable since the connectivity of the line cells does not change under normal circumstances (0-90$^{\circ}$C) \cite{WongRaouxKimEtAl2010}. A bias (yield lower or higher than 50\%) can be removed through simple privacy amplification methods as discussed in Section \ref{subsec:relibility_enhancement} to produce independent and identically distributed bits. 

Annealing the line cells to high temperature will cause void formation at unpredictable regions which may affect cell connectivity; hence adds another degree of randomness. Figure \ref{fig:sem_cell}(c) shows such a case where the void at the center may break the connectivity while the cell looks connected on the surface. 

The cell to cell variations at lithography limit are highly dependent on tools and process parameters; hence, the proposed device is immune to cloning. Also, GST is a soft material compared to silicon, $SiO_2$, $Si_3N_4$ and tends to erode very quickly under focused ion beam. This increases the time, effort and tool complexity required to tamper with the IC. With today's technology the manufacturer cannot choose his own key or find out what key was chosen by the lithography process.

\section{PCM-based POK}
\label{sec:pcm} 

\begin{figure}[!tb]
\centering
\includegraphics[width = 0.8\columnwidth]{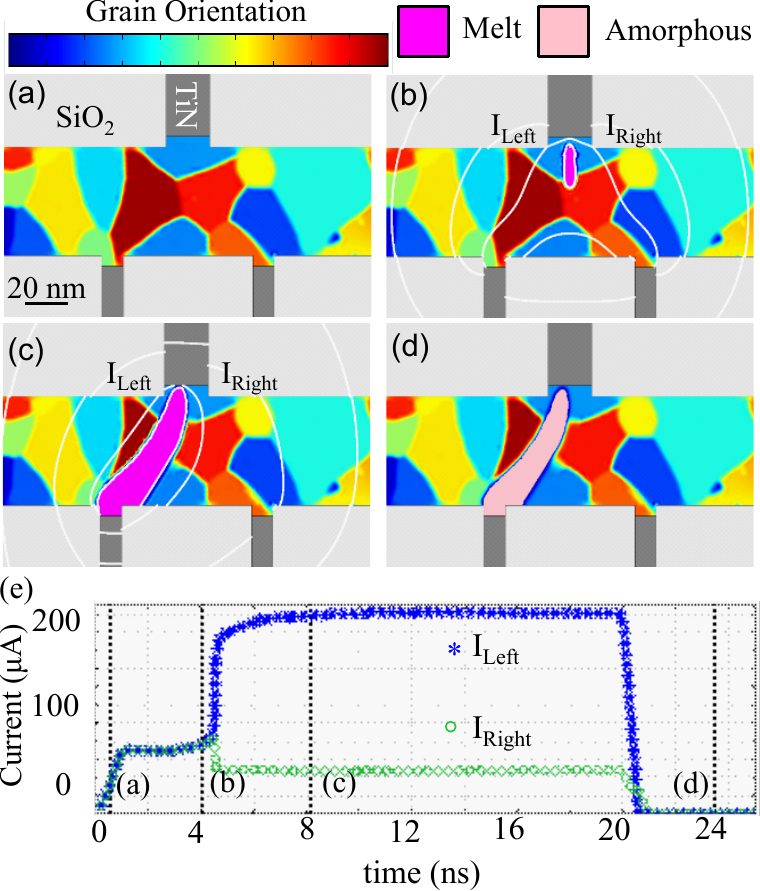}
\caption{{ (a) Polycrystalline GST (rainbow) with three metal contacts (dark gray) and electrically insulating oxide (light gray). (b) Current flows nearly symmetrically from the top contact to the two bottom contacts. (c) When melting begins near one of the bottom contacts, the path to that contact becomes very conductive and the contact draws more of the current. Higher current densities melt more of the path, creating a feedback loop (thermal runaway). (d) A highly resistive amorphous plug blocks the left contact when the current ceases. The initial device-specific grain map determines which contact is plugged, and subsequent read pulses reveal a large resistance contrast between the two paths regardless of drift. (e) Transient current into both contacts.}}
\label{fig:ThreeContactSnapshots}
\end{figure}

GST is considered a nucleation dominated material because over typical crystallization temperatures the amorphous to crystalline transition is primarily driven by new crystal grains nucleating rather than templated growth at existing crystalline/amorphous interfaces. The grain boundary resistivity is higher and displays higher temperature sensitivity than crystalline grain resistivity. The location of grain boundaries depends on stochastic nucleation events and temperature dependent growth velocities during the fabrication process; hence, there is intrinsic device-to-device variability in grain maps. We propose a device that produces a random bit dependent on an initial device specific grain map.  It is not possible to detect grain boundaries or amorphized regions using scanning electron microscopy. This makes a POK using this approach difficult to read with a physical attack. Transmission electron microscopy is capable of such differentiation, but requires time intensive and costly techniques (creating nm scale cross sections and physically moving them into the path of an electron beam), making it difficult to read a large number of bits this way.


Phase change materials typically have negative temperature coefficients of resistivity (TCR) which gives rise to thermal runaway when a sufficiently large voltage pulse is applied. Hence, if there are two competing current paths, one melts and amorphizes while the other does not experience current densities sufficient for melt and remains unchanged. Thermal runaway is determined by asymmetry in the initial grain map (Figure~\ref{fig:ThreeContactSnapshots}a). We demonstrate this concept through PCM device simulations using a finite-element phase change material model~\cite{Dirisaglik2014, Faraclas2014, Faraclas2011} which tracks temperature dependent nucleation, growth, and amorphization in GST (Figure~\ref{fig:ThreeContactSnapshots}). Temperature, phase, and electric-field dependent material parameters for crystalline GST, amorphous GST, and grain boundaries are extracted from experimental measurements~\cite{Adnane2012, Adnane2015, DirisaglikBakanJuradoEtAl2015}. Imperfections in fabrication processes can cause conductivity biases not related to the grain map and should be accounted for in process design.

The high resistivity contrast between amorphous and crystalline states is not affected by small ($\thicksim$ 0.1V) read pulses and crystallization of the amorphous state takes >10 years at 85$^{\circ}$C, making our proposed POK very reliable \cite{WongRaouxKimEtAl2010}.

\section{Conclusion}
\label{sec:conclusion}

We propose two new approaches for intrinsically reliable and lightweight POKs. One approach is based on utilizing the lithography limits to produce line cells that have random functionality. The second approach is a novel PCM-based POK which can achieve a reliable output based on the variability of granularity in PCM cells. Experimental results of lithography-based POK and simulated results of PCM-based POKs are provided.

\section*{Acknowledgment}

The devices are fabricated at IBM T.J. Watson Research Center and characterized at UConn, supported by the U. S. National Science Foundation under awards number ECCS 0925973 and ECCS 1150960. The characterization and analysis efforts of R. S. Khan, N. Noor, J. Scoggin, H. Silva and A. Gokirmak and the problem statement and motivation of C. Jin, H. P. Nguyen and M. van Dijk are supported by AFOSR MURI under award number FA9550-14-1-0351. The authors would like to thank Dr. Chung Lam, Dr. Yu Zhu, Dr. Simone Raoux, Dr. Norma Sosa, Dr. Matthew BrightSky and Dr. Adam Cywar for their contributions to device fabrication at IBM T.J. Watson Research Center.

\newpage
\bibliographystyle{IEEEtran}
\bibliography{puf_detailed_20161110}

\end{document}